\documentstyle[12pt,a41,epsfig]{article}

\newcommand{\ds}{\displaystyle}

\newcommand\GeV{\,\mbox{GeV}}

\newcommand{\bq}{\begin{equation}}
\newcommand{\eq}{\end{equation}}
\newcommand\beq{\begin{equation}}
\newcommand\eeq{\end{equation}}
\newcommand\bea{\begin{eqnarray}}
\newcommand\eea{\end{eqnarray}}

\newcommand\GV{\,\mbox{\boldmath $G$}}
\newcommand\FV{\,\mbox{\boldmath $F$}}

\newcommand\PV{\mbox{\boldmath $P$}}

\newcommand\MV{\,\mbox{\boldmath $M$}}

\newcommand\GA{\,\mbox{\boldmath $\Gamma$}}

\include{epsfig}
\begin{document}
\setlength{\baselineskip}{0.515cm}
\sloppy
\thispagestyle{empty}
\begin{flushleft}
DESY 99--143 \hfill
{\tt hep-ph/9909449}\\
September 1999\\
\end{flushleft}

\mbox{}
\vspace*{\fill}
\begin{center}
{\LARGE\bf QCD Evolution of Structure Functions}

\vspace{2mm}
{\LARGE\bf at Small x}

\vspace{4cm}
\large
Johannes Bl\"umlein\\

\vspace{3mm}
{\it DESY--Zeuthen, Platanenallee 6, D-15735 Zeuthen, Germany}

\end{center}
\normalsize
\vspace{\fill}
\begin{abstract}
\noindent
The status of the resummation of small $x$ contributions to the
unpolarized and polarized deep inelastic structure functions is reviewed.
\end{abstract}

\vspace{1mm}
\noindent

\vspace*{\fill}
\noindent
{\small\sf Lecture notes to appear in the Proceedings of the
Ringberg Workshop "New Trends in HERA Physics 1999", eds.
G. Grindhammer, B. Kniehl, and G. Kramer (Springer, Berlin 1999).}

\newpage
\section{Introduction}
%
The measurement of the nucleon structure functions in deep inelastic
scattering provides important tests of the predictions of Quantum
Chromodynamics (QCD) on the short--distance structure of nucleons. The
experiments at the $ep$--collider HERA allowed to extend the kinematic
region to very small values of $x \sim 10^{-4}$ at photon virtualities
of $Q^2 \geq 10 \GeV^2$ measuring the structure function $F_2(x,Q^2)$
at an accuracy of $O(1\%)$. These precise measurements allow dedicated
tests of QCD. Also the polarized deep inelastic experiments approach
smaller values of $x$ with a higher accuracy.
To obtain a description in this kinematic range potentially large
contributions to the evolution kernels were studied during the last two 
decades and the resummations of `leading term' contributions were
performed. Here mainly two directions were followed.

In one approach \cite{GLR} a non--linear resummation of fan--diagrams
of single ladder cascades is performed
in the double logarithmic 
approximation~\cite{DLA}\footnote{This approximation has to be
considered as qualitative and leads often to an overestimate of the
scaling violations, cf.~\cite{BV5}.}. 
Corrections of this type may ultimately become
important at very small values of $x$ to restore
unitarity. Numerical 
studies of this equation were performed in Refs.~\cite{GLR1}. 
One important
assumption in the solution of this equation was that the 
nonperturbative input distributions for the $N$--ladder terms are given
by the Nth power of the single gluon distribution at some starting
scale. This would imply a strong constraint on the hierarchy of higher
twist distributions. Later it was found \cite{BAR} that the approximation
\cite{GLR} has to be supplemented by further color correlations even
in the double logarithmic approximation, which cannot be cast into a
non--linear equation  anymore. As implied by the operator
product expansion, the contributions due to different twist renormalize
independently. The corresponding input
distributions are likely to be unrelated between the different twists.
Still saturation effects of the structure functions at very small $x$
may be caused due to higher twist contributions. However, the detailed
dynamics is yet  unknown.

Ladder equations also form the basis of other approaches. In a physical
gauge the emission of gluons along a single ladder--cascade 
describes\footnote{The virtual contributions have to be added.} in 
{\it leading order} (LO) the evolution of a parton density as predicted
by the renormalization group equation if the emissions along the
ladder are strongly ordered in the transverse momentum $k_{\perp, 1}
\ll ... k_{\perp,i} \ll k_{\perp,i+1} ...$ If these emissions are
evaluated in the approximation $x_1 \gg ... x_i \gg x_{i+1} ...$ 
instead, using effective vertices, one obtains the 
BFKL--resummation~\cite{BFKL} in LO. This particular aspect led sometimes
to the impression that these two approximations were of competing
nature. As we will show below this is, however, not the case as
far as the description of the scaling violations of structure functions
are concerned.
One may study this process under a more general point of
view and consider 
angularly ordered emissions covering both the above cases~\cite{MAR}, 
which allows for interesting applications through Monte Carlo studies.
Whereas this unified treatment is possible at LO, higher order 
corrections cannot be cast into this form in general. The 
renormalization group equation for the mass singularities, on the other
hand, allows to
perform consistent higher order calculations beyond these approximations
accounting for the resummation of the small $x$  contributions in
the anomalous dimensions and coefficient functions.

In the second main approach these resummations are studied.
Resummations were performed
in leading order for the unpolarized singlet
case~\cite{BFKL,JAR}, the non--singlet structure functions~\cite{KL,BV1},
and the polarized singlet distributions~\cite{BER2}. Applications
were studied in the case of QED for the flavor non--singlet 
contributions to radiative corrections~\cite{BV2,BVR2}.
The quarkonic
next--to--leading order (NLO) contributions in the unpolarized singlet 
case were calculated in \cite{CH}. Recently also the NLO resummed
gluon anomalous
dimension~\cite{FL,CC2} in the DIS-$Q_0$ scheme~\cite{CIA} was obtained.
If the evolution kernels are written in terms of a series in 
$\alpha_s/(N-N_s)$, where $N_s$ denotes the position of the leading pole,
the individual terms are large and require resummation.

One of the central questions for the understanding of the deep inelastic
structure functions at small $x$ is therefore to analyse the impact and
r\^{o}le of these small $x$
resummations and their potential corrections
in even higher order. These terms have to be viewed in comparison with
the known fixed order results used in the current analyses of the
scaling violations of the twist--2 contributions to the structure
functions.

In the present paper we review the status of the latter resummations
and their impact on the scaling violations of deep inelastic structure
functions.
\section{The Evolution Equations}
%
The twist-2 contributions to the structure functions in inclusive
deep-inelastic scattering can be described in terms of the QCD-improved
parton model. Their scaling violations are governed by renormalization
group equations which can be formulated to all orders in the strong
coupling constant. All small $x$ resummations are based on perturbative
QCD. As in the fixed order calculations one has to factorize the 
collinear or mass singularities, which are absorbed into the 
non--perturbative input distributions. The soft- and virtual 
singularities cancel order by order according to the Bloch--Nordsiek 
theorem. A second renormalization group equation describes the scale 
dependence of the strong coupling constant $\alpha_s(\mu^2)$.
The perturbative all-order 
small $x$ resummations may turn out to yield important contributions to 
the {\it scaling violation} of the deep inelastic structure functions.
Predictions on the shape of the parton densities at
small $x$ are, however, {\it beyond} a perturbative treatment, even in
resummed form, since generally low scales are involved and partonic
approaches have to fail.

The small $x$ resummations can be tested with respect to their
prediction on the scaling violations of deep inelastic structure 
functions as $F_2(x,Q^2)$ and $F_L(x,Q^2)$. The evolution equations
for the parton densities $f_i(x,\mu^2)$ are given by
\begin{equation}
\frac{\partial}{\partial \log(\mu^2)} f_i(x,\mu^2) =  P_i^j(x,a_s)
\otimes f_j(x,\mu^2)~.
\end{equation}
Here
$\otimes$ denotes the Mellin convolution.
The splitting functions $P_{ij}(x,a_s)$
contain besides the completely known
LO and NLO contributions the LO and NLO small $x$ resummed terms to
all orders in $a_s=\alpha_s/(4\pi)$,
\begin{equation}
P_{ij}(x,a_s) = a_s P_{ij}^{(0)}(x) + a_s^2 P_{ij}^{(1)}(x)
+ 
\sum_{k=2}^{\infty} a_s^{k+1} \widehat{P}_{ij,x \rightarrow 0}^{(k)}(x)
+ 
\sum_{k=2}^{\infty} a_s^{k+2} \widehat{P}_{ij,x 
\rightarrow 0}^{(k)}(x)~.
\end{equation}
Similarly, the coefficient functions take the form
\begin{eqnarray}
c_{i}\left(x, \frac{Q^2}{\mu^2}\right) &=&
 \delta_{iq} \delta(1-x)
a_s c_{i}^{(0)}(x) + a_s^2 c_{i}^{(1)}(x) \nonumber\\
& &
+ 
\sum_{k=2}^{\infty} a_s^{k+1} \widehat{c}_{i,x \rightarrow 0}^{(k)}(x)
+ 
\sum_{k=2}^{\infty} a_s^{k+2} \widehat{c}_{i,x \rightarrow 0}^{(k)}(x)~.
\end{eqnarray}
In this way the effect of the small $x$ resummations is consistently
included. As these contributions do not a priori account for
Fermion number and energy--momentum conservation these conditions
have to be imposed for the contributions beyond $O(a_s^2)$.
The structure functions $F_A(x,Q^2)$ are finally obtained as
\begin{equation}
F_A(x,Q^2) =c_{q,A}\left(x,\frac{Q^2}{\mu^2}\right) \otimes f_q\left(
x,\frac{\mu^2}{M^2}\right)
           +c_{g,A}\left(x,\frac{Q^2}{\mu^2}\right) \otimes f_g\left(
x,\frac{\mu^2}{M^2}\right)~.
\end{equation}
The factorization scale dependence ($\mu^2$) cancels order by order.
\section{Small x Resummation of the Anomalous Dimensions}
%
All resummations studied below are based on scale--invariant equations
in leading order. If one considers the renormalization group equation
for an operator matrix element $E_k^n$
\begin{equation}
\left[\mu \frac{\partial}{\partial \mu} + \beta \frac{\partial}
{\partial g} + \gamma_m m \frac{\partial}{\partial m} + \gamma_{O_k}
- n \gamma_{\Phi} \right] E_k^n = 0
\end{equation}
scale invariant solutions are obtained in the massless case ($m=0$) and
iff the $\beta$--function is set to zero~:
\begin{equation}
E_k^n(\mu^2) = E_k^n(\mu_0^2) \left(\frac{\mu^2}{\mu_0^2}\right)^
{\frac{1}{2}\left(\gamma_{O_k} - n \gamma_{\Phi}\right)}~.
\label{eqCONF}
\end{equation}
Within this approach the coupling constant $a_s$ is fixed. The scale
invariant part of the anomalous dimension has the representation
\begin{equation}
\gamma_{O_k} - n \gamma_{\Phi} = \sum_{l=1}^{\infty} 
\gamma_O^{(l)} a_s^l
\end{equation}
and exponentiates to all orders. The representation (\ref{eqCONF}) 
applies also for higher order resummations under the above requirements.
In this way one may derive in the different subsequent resummations
the LO small $x$ resummed anomalous dimensions. In higher than LO
scale breaking effects emerge in QCD. Therefore a thorough treatment
along these lines is no longer possible. Still one may try to identify
those contributions of the anomalous dimension which are scale invariant
applying a diagonalization as in (\ref{eqCONF}).
\section{Less Singular Terms}
%
For most of the applications only the resummation of the leading
singular terms is known. The contributions which are less singular
by one or more powers in $N$ may yield substantial contributions.
This  has been known for long~\cite{DURH,WIL},
cf. also \cite{MW}, and 
can easily be seen
in the case of $F_L(x,Q^2)$ in $O(\alpha_s)$ as an example. If one
disregards the second factor in the leading order coefficient function
$c_g^{(0)} = C_g^{(0)} x^2(1-x)$ in view of a small $x$ approximation
the value of $F_L$ may be overestimated by a factor of four~\cite{DURH}.

\begin{figure}[h]
\begin{center}
\includegraphics[width=0.6\textwidth]{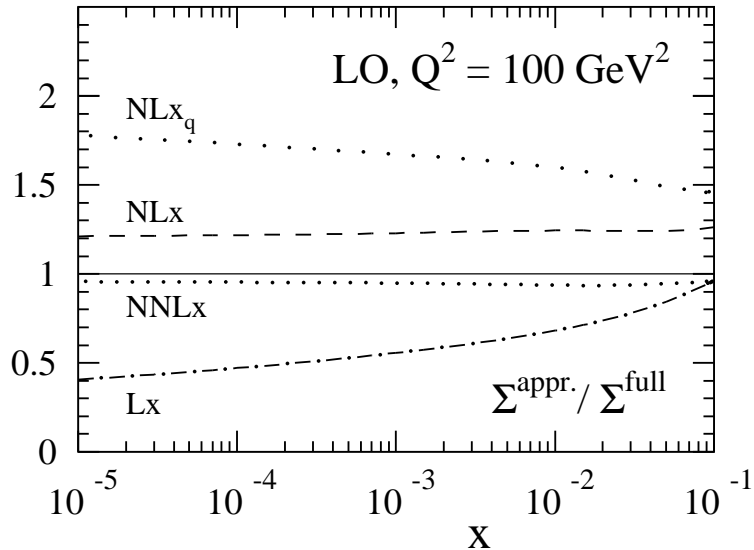}
\end{center}

\vspace{-7mm}
\caption[]{Different approximation steps in the $1/(N-1)$ expansion
of the complete LO unpolarized singlet distribution over four orders
(Lx, NLx$_q$, NLX, NNLX), cf. Ref.~\cite{BV5}.}
\label{BL-figsub}
\end{figure}

To get an estimate of the effect of the
terms suppressed by one order or more orders in the Mellin moment $N$
one may study some models. The possible size of these terms may be 
inferred expanding the LO and NLO anomalous dimensions and coefficient
functions into series in $1/N$ comparing   the expansion coefficients.
Estimates of this kind were performed in
[3,10,12,13,21--25].
Possible ans\"atze for the next order terms are
\begin{eqnarray}
\Gamma(N,a_s) &\rightarrow& \Gamma(N,a_s) - \Gamma(1,a_s) \nonumber\\
\Gamma(N,a_s) &\rightarrow& \Gamma(N,a_s)(1-N)            \nonumber\\
\Gamma(N,a_s) &\rightarrow& \Gamma(N,a_s)(1-N)^2          \nonumber\\
\Gamma(N,a_s) &\rightarrow& \Gamma(N,a_s)(1-2N+N^3)~.
\end{eqnarray}
If one formally expands the LO and NLO anomalous dimensions in the
above manner one finds~\cite{BV5}, irrespectively of the factorization
scheme,  that at least four expansion terms are needed to represent
the exact result on the $5\%$ level, cf.~figure~\ref{BL-figsub}.
If one compares the respective NLO resummed coefficients with the LO
resummed ones in the cases they were calculated even larger effects
than indicated by the above estimate are found (see below).
\section{Non--Singlet Structure Functions}
%
The most singular contributions to the Mellin transforms of the 
structure--function evolution kernels $K^{\pm}(x,a)$ at all orders in 
$a$ can be obtained from the positive and negative signature amplitudes 
$f_0^{\pm}(N, a)$ studied in \cite{KL} for QCD via
\begin{equation}
{\cal M} \left [ K_{x \rightarrow 0}^{\pm}(a) \right ](N) 
 \equiv \int_0^1 \! dx \, x^{N-1} K_{x \rightarrow 0}^{\pm}(x, a)
 \equiv - \frac{1}{2} \GA_{x \rightarrow 0}^{\pm}(N, a)
 = \frac{1}{8\pi ^2} f_0^{\pm}(N, a) \: .
\end{equation}
These amplitudes are subject to the quadratic equations:
\begin{eqnarray}
f_0^+(N, a) &=&  16 \pi^2 a_0  \frac{a}{N} + \frac{1}{8 \pi^2}
 \frac{1}{N} \left[ f_0^+(N, a) \right] ^2 \: , \label{eqf0p} \\
f_0^-(N, a) &=&  16 \pi^2 a_0  \frac{a}{N} + 8 b_0^-  \frac{a}{N^2}
 f_V^+(N, a) + \frac{1}{8 \pi^2} \frac{1}{N} \left [ f_0^-(N, a)
 \right ]^2 \: .  \label{eqf0m}
\end{eqnarray}
Here $f_V^+(N, a)$ is obtained as the solution of the Riccati 
differential equation
\begin{equation}
\label{eqRIC}
 f_V^+(N, a) = 16 \pi^2 a_V  \frac{a}{N} + 2 b_V  \frac{a}{N}
 \frac{d}{d N} f_V^+(N, a) + \frac{1}{8 \pi^2} \frac{1}{N}
 \left [f_V^+(N, a)\right]^2 \:\: .
\end{equation}
The coefficients $a_i$ and $b_i$ in the above relations read for the 
case of QED, cf. sect.~7,
\begin{equation}
 a_0 = 1,~~b_0^- = 1,~~a_V = 1,~~b_V = 0,
\end{equation}
and for QCD~\cite{KL}
\begin{equation}
 a_0 = C_F,~~b_0^- = C_F,~~a_V = -\frac{\ds 1}{2 N_c},~~b_V 
 = C_A,
\end{equation}
with $C_F = 4/3$ and $C_A = N_c = 3$.
In QED Eq.~(\ref{eqRIC}) further simplifies to an algebraic equation
with the same coefficients as (\ref{eqf0p}). The solutions of
(\ref{eqf0p}) and (\ref{eqf0m}) were derived in \cite{KL} for the 
QCD case\footnote{Note a few misprints in Eq.~(4.7) of
ref.~\cite{KL}.}. They are given by 
\begin{eqnarray}
 \GA_{x \rightarrow 0}^{+}(N, a) &=& - N
  \left \{ 1 - \sqrt{1 - \frac{8 a C_F}{N^2}} \right \} \\
 \GA_{x \rightarrow 0}^{-}(N, a) &=& - N
  \left \{ 1 - \sqrt{1 - \frac{8 a C_F}{N^2}
  \left [1 - \frac{8 a N_c}{N} \frac{d}{d N}
  \ln \left ( e^{z^2/4} D_{-1/[2N_c^2]}(z) \right ) \right ] } 
  \right \}  \nonumber
\label{eqGAA}
\end{eqnarray}
where $z = N/\sqrt{2 N_c a}$, and $D_p(z)$ denotes the function of the 
parabolic cylinder.

\begin{figure}[h]
\begin{center}
\includegraphics[width=.5\textwidth]{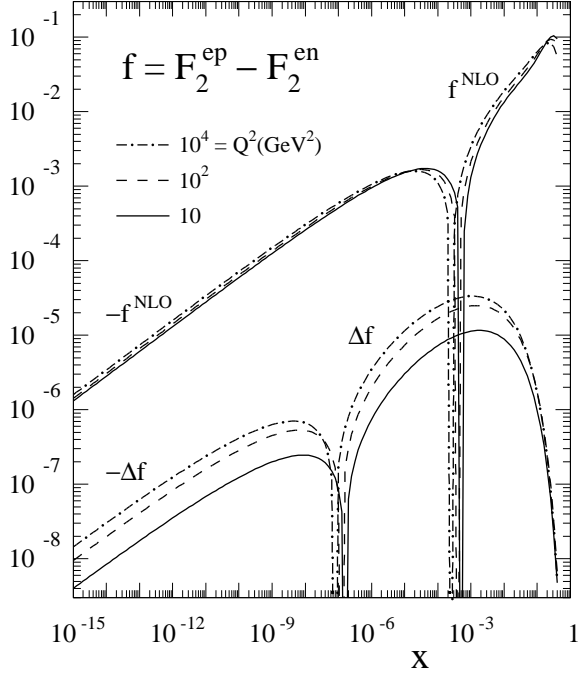}
\end{center}

\caption[]{The small-$x$ $Q^2$--evolution of the unpolarized 
non--singlet structure function combination 
$F_2^{\, ep} - F_2^{\, en}$ 
in NLO and the absolute corrections to these results due to the 
resummed kernel. The initial
distributions were chosen
at $ Q_{0}^{2} = 4 \mbox{ GeV}^2 $, 
cf.~Refs.~\cite{BV1,BV2}.} 
\label{BL-figf2NS}
\end{figure}

Numerical results on the impact of the small $x$ resummations were
obtained in Refs.~\cite{BV1,BV2}. The new terms contribute at $O(a_s^3)$
and higher. If compared to the fixed order contributions in NLO the 
effect is of $O(1\%)$ or less, which is shown for the scaling  violations
of $F_2^{ep}-F_2^{en}$ in 
Fig.~\ref{BL-figf2NS}. Other examples as for
$xF_3$
and the $\pm$--evolutions for polarized non--singlet structure functions
show a rather similar behaviour. Comparable results were obtained in
Ref.~\cite{KOD}. The effect is expected to be rather small due to
the typical shape of the input distributions in the non--singlet case.
The size of the (small) correction does further vary
significantly in dependence of the inclusion of less singular 
terms, cf. sect.~4, or if conservation laws are imposed. Large effects
as anticipated in Refs.~\cite{EMR,BER1} are not confirmed.
\section{Polarized Singlet Structure Functions}
%
The LO small $x$ evolution kernels in the case of the polarized singlet
evolution were derived in \cite{BER2}. The resummed splitting
function is given by
\begin{equation}
\PV(x,a_s) \equiv \sum_{l=0}^{\infty} \PV^{(l)}_{x\rightarrow 0} 
a_s^{l+1} \log^{2l} x = \frac{1}{8\pi^2} {\cal M}^{-1}[\FV_0(N,a_s)](x).
\label{eqSIPO}
\end{equation}
The matrix valued function $\FV_0(N,a_s)$ is obtained as the
solution of 
\begin{equation}
\FV_0(N,a_s) = 16 \pi^2 \frac{a_s}{N} \MV_0 - \frac{8a_s}{N^2} 
\FV_8(N,a_s) \GV_0
+ \frac{1}{8\pi^2}\frac{1}{N} \FV_0^2(N,a_s)
\end{equation}
with
\begin{equation}
\FV_8(N,a_s) = 16 \pi^2 \frac{a_s}{N} \MV_8 + \frac{2a_s}{N} C_A 
\frac{d}{dN} \FV_8(N,a_s) + \frac{1}{8\pi^2} \frac{1}{N} \FV_8^2(N,a_s),
\end{equation}
where
\begin{equation}
\MV_0 = \left(\begin{array}{cc} C_F & -2T_R N_f \\ 2C_F & 4C_A
\end{array} \right),~
\MV_8 = \left(\begin{array}{cc} C_F - C_A/2 & -T_R N_f \\ C_A & 2 C_A
\end{array} \right),~
\GV_0 = \left(\begin{array}{cc} C_F & 0 \\ 0 & C_A
\end{array} \right).
\end{equation}
\begin{figure}[h]
\begin{center}
\includegraphics[width=.9\textwidth]{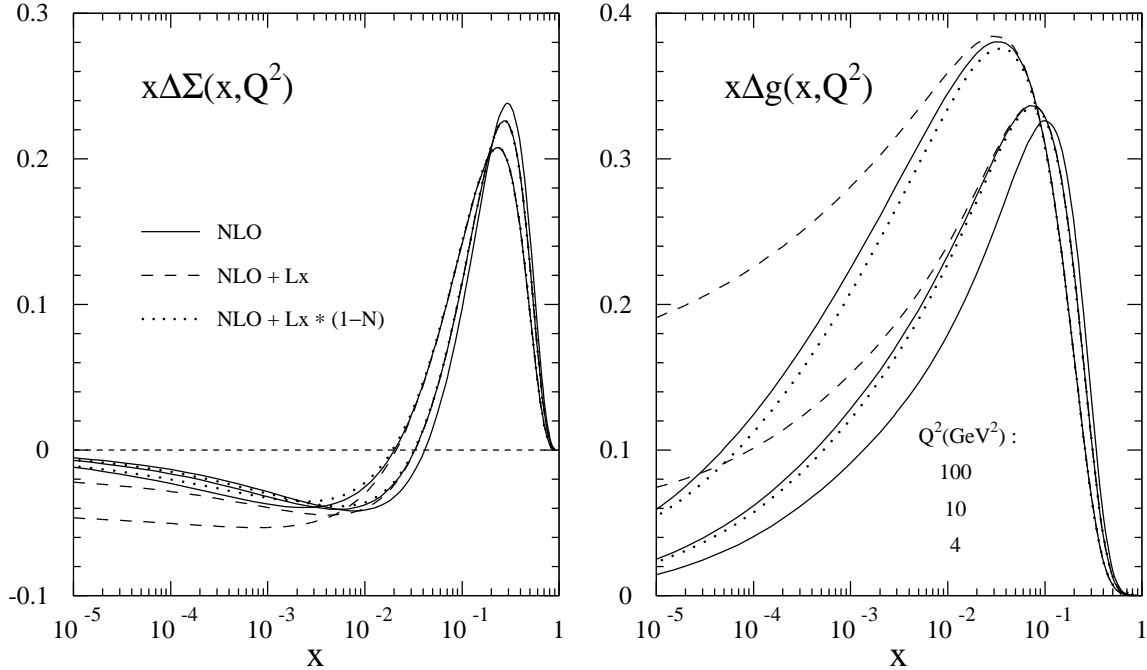}
\end{center}

\caption[]{The $Q^2$ evolution of the polarized quark singlet and gluon
momentum distributions evolving from 
$Q_0^2 = 4 \GeV^2$,~Ref.~\cite{BV3}.}
\label{BL-figG1}
\end{figure}
Eq.~(\ref{eqSIPO}) obeys~\cite{BV3}
\begin{equation}
P_{qg}^{(l)}/(T_R N_f) = - P_{gq}^{(l)}/C_F
\end{equation}
to all orders, where $T_R=1/2$ and $N_f$ denotes the number of flavors.
The leading contributions of the fixed order results
in LO and NLO ($\overline{\rm MS}$) are correctly described. In the
supersymmetric limit $C_A=C_F=N_f=1, T_R=1/2$ the relations
\begin{equation}
P_{qq}^{(l)}+P_{gq}^{(l)} = P_{qg}^{(l)}+P_{gg}^{(l)}
\end{equation}
are
obeyed for all $l$ and Eq.~(\ref{eqSIPO}) can be given in a simple
analytic form~\cite{BV3}.

The impact of the resummation (\ref{eqSIPO}) on the evolution of the
polarized singlet and gluon density and the structure function 
$g_1(x,Q^2)$ have been studied in Ref.~\cite{BV3}. As shown in
Fig.~\ref{BL-figG1} the corrections are much larger than the
$O((a_s \ln^2 x)^l)$ corrections in the non--singlet case. Taking 
into account less singular terms of the type
\begin{equation}
P_{ij}^{(l>1)} \rightarrow
P_{ij}^{(l>1)} \cdot (1-N)~,
\end{equation}
as suggested by the analytic structure of the fixed order LO and NLO
anomalous dimensions, this enhancement reduces, however,
again to the value
of the fixed order evolution in NLO (dotted line in Fig.~\ref{BL-figG1}).
\section{QED Corrections}
\begin{figure}[h]
\begin{center}
\includegraphics[width=0.6\textwidth]{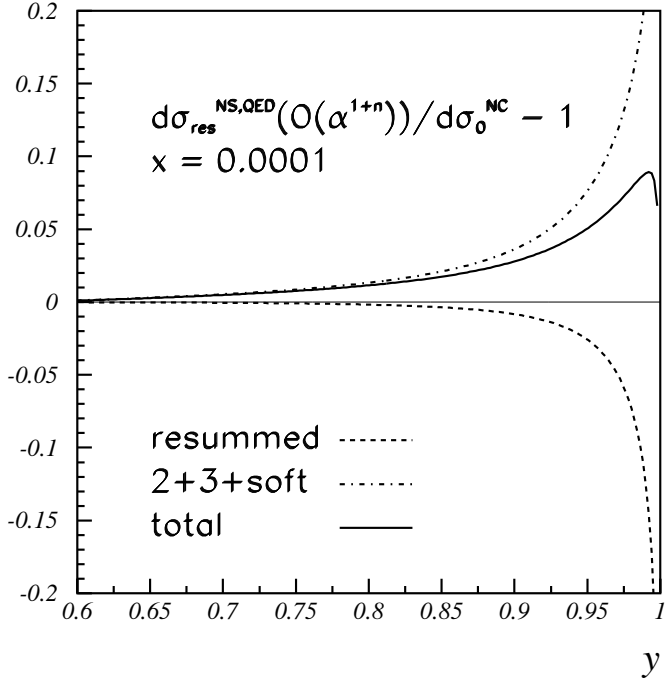}
\end{center}
\caption[]{2nd and higher order QED
initial state radiative corrections
to deep inelastic $ep$ scattering. Dashed line: small $x$ resummed
contribution, dash-dotted line: LO contributions up to $O(\alpha^3)$
and soft photon exponentiation, full line: resulting correction,
Ref.~\cite{BVR2}.}
\label{BL-figQED}
\end{figure}

%
The non--singlet small $x$ resummation was applied to resum the
$O((\alpha \log^{2} x)^l)$ 
terms in the QED corrections to deep
inelastic scattering in Ref.~\cite{BVR2}. These corrections are negative
and amount to $O(10\%)$ in the high $y$ range for 
$x= 10^{-4} ... 10^{-2}$, see Fig.~\ref{BL-figQED}.

They diminish the
$O((\alpha \log(Q^2/m_e^2))^l)$ corrections which are very large in this
domain. The first non-trivial contribution of 
$O(\alpha^2 \log^{2} x)$ is in agreement with the result found
in~\cite{BBN}. From the latter calculation also the next less singular 
term of $O(\alpha^2 \log(x))$ can be derived. Up to this term the
evolution kernel reads ($a = \alpha/(4\pi)$)
\begin{equation}
{\cal M}[P_{z \rightarrow 0}](N,a) = \frac{2a}{N} - 12 \frac{a^2}{N^3}
\left(1- \frac{2}{9} N\right) + ...
\end{equation}
If compared to the case of QCD this less--singular term is stronger 
suppressed.
\section{Unpolarized Singlet Distributions}
%
The LO resummation for the evolution kernel of the
 unpolarized singlet
distributions was derived in \cite{BFKL}. Jaroszewicz~\cite{JAR} showed
that the eigenvalue 
\begin{equation}
(N-1)       = \frac{\alpha_s N_c}{\pi} \chi_0(\gamma_L)
      \equiv
\frac{\alpha_s N_c}{\pi} \left[2 \psi(1) - \psi(\gamma_L) 
- \psi(1-\gamma_L)\right]
\label{eq1BF}
\end{equation}
represents the LO resummed gluon-gluon anomalous dimension 
$\gamma_L = \gamma_{gg}^{(0)}(N,a_s)$. The resummed LO
gluon-quark anomalous dimension is given by $\gamma_{gq}^{(0)}(N,a_s)
= (C_F/C_A) \gamma_L$ and the quarkonic terms do not contribute in
$O((a_s/(N-1))^l$. Eq.~(\ref{eq1BF}) can be solved iteratively
demanding $\gamma_L(N,a_s) \rightarrow \overline{\alpha}_s/(N-1)$ as
$|N| \rightarrow \infty$ for $N~\epsilon~{\rm\bf C}$, which selects
the physical branch of the resummed anomalous dimension,
\begin{equation}
\gamma_L \equiv \gamma_{gg,0}(N,\alpha_s) =
\frac{\overline{\alpha}_s}{N-1} \left\{1 + 2 \sum_{l=1}^{\infty} 
\zeta_{2l+1} \gamma_{gg,0}^{2l+1}(N,\alpha_s) \right\}.
\end{equation}
Here we rewrite
$\overline{\alpha}_s = N_c \alpha_s/\pi$. $\gamma_L$
has the serial representation
\begin{equation}
\gamma_{gg,0}(N,\alpha_s) = \frac{\overline{\alpha}_s}{N-1}
+ 2 \zeta_3 \left(\frac{\overline{\alpha}_s}{N-1}\right)^4
+ 2 \zeta_5 \left(\frac{\overline{\alpha}_s}{N-1}\right)^6
+ 12 \zeta_3^2 \left(\frac{\overline{\alpha}_s}{N-1}\right)^7
+ \ldots
\label{eqPOLE}
\end{equation}
Under the above conditions one may calculate $\gamma_L(N,a_s)$ in the
whole complex plane. It is a bounded function of
$\rho = (N-1)/\overline{\alpha}_s$, the singularities of which are
branch points at~\cite{JB95,EHW}
\begin{equation}
\rho_1 = 4 \log 2,~~~~~\rho_{2,3} = -1.41048 \pm 1.97212~i,
\end{equation}
cf.~\cite{JB95,BV5} for  detailed representations.
The LO BFKL anomalous dimension possesses {\bf no poles}.
Since the known NLO resummed anomalous dimensions  are functions of 
$\gamma_L(N,a_s)$ which introduce no further singularities the contour
integral around the singularities of the problem has to cover the
three BFKL branch points, the singularities of the input distributions
along the real axis left of 1, and the remaining singularities of the
fixed order anomalous dimensions at the non--positive 
integers~\cite{JB95,BV5}. Note, that the resummed form of 
$\gamma_L(N,a_s)$ removes {\it all} the fixed--order
pole singularities of
Eq.~(\ref{eqPOLE}) into branch cuts. Any finite correction to $\gamma_L$
may thus lead to essential changes of the corresponding numerical
results. Early numerical studies on the impact of the LO resummed
anomalous dimensions were performed in \cite{KWI}. More recent
analyses have been performed in Refs.~\cite{BV5,EHW,BVR1}.

The next-to-leading order resummed anomalous dimensions are given by
\renewcommand{\arraystretch}{1.3}
\begin{equation}
\widehat{\gamma}_{NL}(N,\alpha_s) =  -2 \left(
\begin{array}{ll} {\ds \frac{C_F}{C_A}} \left[\gamma_{qg}^{NL} -
{\ds \frac{8}{3}} a_s T_F \right] & \gamma_{qg}^{NL} \\
\gamma_{gq}^{NL} &  \gamma_{gg}^{NL} \end{array}\right)~,
\end{equation}
\renewcommand{\arraystretch}{1.}
with $T_F = T_R N_f$.
The quarkonic contributions were calculated in Ref.~\cite{CH}, as well
as the resummed coefficient functions $c_2(N,a_s)$ and $c_L(N,a_s)$.
Recently $\gamma_{gg}^{NL}$ was derived in \cite{FL,CC2} and
$\gamma_{gq}^{NL}$ is yet unknown\footnote{As pointed out in 
Ref.~\cite{BV5} its quantitative influence is likely to be minor.}. 
In the DIS--scheme $\gamma_{qg}^{NL}$ is found to be an
analytic, scale--independent
function of $\gamma_L(N,a_s)$ and reads
\begin{equation}
\gamma_{qg}^{NL,DIS}(N,\alpha_s) = T_F \frac{\alpha_s}{6\pi}
\frac{2 + 3 \gamma_L - 3 \gamma_L^2}{3-2\gamma_L} 
\frac{\left[B(1-\gamma_l,1+\gamma_L)\right]^3}
{B(2+2\gamma_L,2-2\gamma_L)} R(\gamma_L),
\end{equation}
where
\begin{equation}
R(\gamma) = \left[\frac{\Gamma(1-\gamma) \chi_0(\gamma)}{- \gamma
\Gamma(1+\gamma) \chi_0'(\gamma)}\right]^{1/2} \exp\left[\gamma \psi(1)
+
\int_0^\gamma dz \frac{\psi'(1)-\psi'(1-z)}{\chi_0(z)}\right]~.
\end{equation}

The NLO resummed gluon anomalous dimension $\gamma_{gg}^{NL}$ was
calculated in the $Q_0$--scheme\footnote{For a transformation into the
DIS--scheme cf. \cite{BV5}.}.
One has to solve the Bethe--Salpeter equation
\begin{equation}
(N-1) G_N(q_1,q_2) = \delta^{D-2}(q_1-q_2) + \int d^{D-2} q_3 K(q_1,q_2)
G_N(q_3,q_2)
\end{equation}
with
\begin{equation}
K(q_1,q_2) = \delta^{D-2}(q_1 - q_2) 2 \omega(q_1) 
+ K_{\rm real}(q_1,q_2)
+ K_{\rm virtual}(q_1,q_2).
\end{equation}
For
$q_1^2 \gg q_2^2$ one diagonalizes as in the LO case using {\it formally}
the same ansatz~:
\begin{equation}
\int d^{D-2} d q_2 K(q_1,q_2) \left(q_2^2\right)^{\gamma-1}
= \overline{\alpha}_s \left[\chi_0(\gamma) 
- \frac{\overline{\alpha}_s}{4} \delta(\gamma,q_1^2,\mu^2)\right]
\left(q_1^2\right)^{\gamma-1}~.
\end{equation}
Here the scale--invariant
LO eigenvalue $\overline{\alpha}_s  \chi_0(\gamma)$ is
supplemented by the NLO correction term $(\overline{\alpha}^2_s/4)
\delta(\gamma, q_1^2,\mu^2)$,
\begin{eqnarray}
\label{eqDEL}
\delta(\gamma,q_1^2,\mu^2) &=&
 - \left(\frac{67}{9} - 2 \zeta(2) -
\frac{10}{27} N_f \right) \chi_0(\gamma) 
+ 4 \Phi(\gamma) - \frac{\pi^3}
{\sin^2(\pi \gamma)} \nonumber\\ & &
+ \frac{\pi^2}{\sin^2(\pi\gamma)} \frac{\cos(\pi
\gamma)}{1- 2 \gamma} \left[(22-\beta_0)
+ \frac{\gamma(1-\gamma)}{(1+2 \gamma) (3- 2 \gamma)} \left(1 +
\frac{N_f}{3}\right)\right]
\nonumber\\
& & +
\frac{\beta_0}{3} \chi_0(\gamma) \log\left(\frac{q_1^2}{\mu^2}
\right)
+ \left[\frac{\beta_0}{6} + \frac{d}{d\gamma}\right]
\left[\chi_0^2(\gamma) + \chi_0'(\gamma)\right] - 6 \zeta_3
,
\end{eqnarray}
with
\begin{equation}
\Phi(\gamma) = \int_0^1 \frac{dz}{1+z} \left[z^{\gamma-1} + z^{\gamma}
\right] \left[{\rm Li}(1) - {\rm Li}(z)\right]~.
\end{equation}
Whereas the contributions in the first two lines of Eq.~(\ref{eqDEL})
do      contain contributions to the anomalous dimension up to $O(a_s^2)$
the third line contributes only with three--loop order. The former terms
are {\it
scale--invariant}
 and are in agreement with the known fixed order 
results. Eq.~(\ref{eqDEL}) therefore makes a prediction on the small $x$
contributions of the yet unknown gluon anomalous dimension in 
three--loop and higher order, which will 
be tested in the future. Besides the scale--dependent term $(\beta_0/3)
\chi_0(\gamma) \log(Q^2/\mu^2)$ also the second addend depends on the
choice of scales, since it is not invariant against the interchange
of $q_1^2$ and $q_2^2$, cf.~\cite{CC2}. The third addend $6 \zeta_3$,
being numerically large, contains contributions of the gluonic 
contribution to the trajectory 
function $\omega(q_1^2)$. The result given in Ref.~\cite{FAD1} was
confirmed in a different calculation by Ref.~\cite{BNR}. A departing
value was reported in \cite{KOR}.

\begin{figure}[h]
\begin{center}
\includegraphics[width=0.4\textwidth]{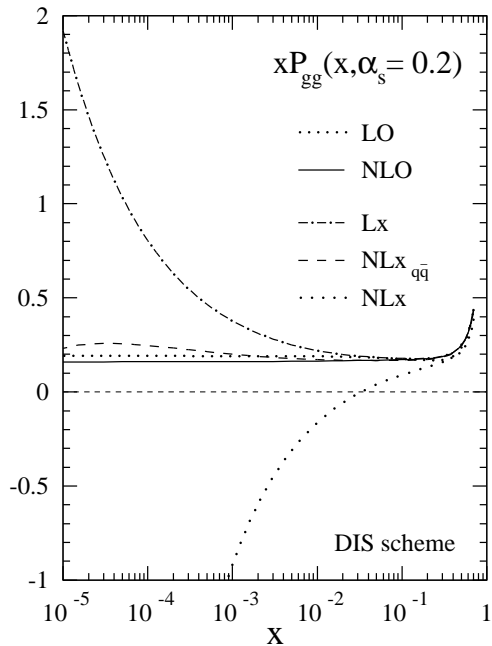}
\end{center}
\caption[]{Different contributions to the resummed splitting function
$xP_{gg}(x,\alpha_s)$ in the DIS--scheme (overlayed),~Ref.~\cite{BRNV}.}
\label{BL-figPGG}
\end{figure}


Numerical results on the impact of the leading and next--to--leading
anomalous dimensions and coefficient functions were provided in a series
of detailed studies, see e.g.~\cite{EHW,BV5,BRNV} and references 
therein. The matrix formalism for the solution of the all order evolution
equations, extending a first approach in Ref.~\cite{EKL}  to all orders,
both for hadronic and photon structure functions, is described
in Ref.~\cite{BV5} in detail.
The quarkonic contributions lead to a strong enhancement of both
$F_2(x,Q^2)$ and $F_L(x,Q^2)$ at small $x$ during the evolution.
However, already simple choices for the yet unknown less singular 
contributions diminish these effects sizably so that a final conclusion
cannot be drawn at present. In the case of the resummed gluon anomalous 
dimension the NLO contributions are found to be extremely large and
negative. The large rise due to the LO BFKL term is already canceled
to the level of the fixed order contributions by the purely
quarkonic contribution to $\gamma_{gg}^{NL}$, see Fig.~\ref{BL-figPGG}.
Adding also the gluonic contribution leads to negative values for
the resummed splitting function already for $\alpha_s = 0.2$ and
$x \simeq 0.01$ which has  to be regarded as unphysical. The LO and NLO
resummed contributions to the gluon anomalous dimension seem to represent
the first terms of a diverging series, which might be eventually resummed.
This can, however, only be achieved reliably if several
more less singular series are calculated  {\it completely}, but not at
the present stage.
\section{An Exactly Soluble Model}
%
The effect of potential subleading contributions to the LO anomalous 
dimension was estimated
in the previous sections in the case of QCD. In
$\phi^3$ theory in $D=6$ dimensions these terms can be determined 
in explicit form. $\phi^3_6$ theory is rather similar to QCD 
(gluo--dynamics)  due to the triple boson interaction and being
 an asymptotic 
free field theory. The leading order resummed anomalous dimension can
be calculated for {\it all} values of $x$ solving the Bethe--Salpeter
equation~\cite{LOV}
\begin{equation}
T(p,q) = \frac{2^{2-D}}
{\pi^{D/2} \Gamma((D-2)/2)} \frac{\lambda^2_D}{(p-q)^2}
+ \frac{\lambda_D^2}{(2\pi)^D} \int d^Dk \frac{T(k,p)}{(q-k)^2 \left(k^2
\right)^2}~,
\label{eqBS}
\end{equation}
with $q^2, p^2 < 0$, $q$ the momentum transfer and $p^2$ a hadronic
mass scale.
For $D=6$ the quantity $q.p~T(p,q)$ is scale invariant and one may expand
Eq.~(\ref{eqBS}) into partial waves with
\begin{equation}
p.q~T_N(p,q) = \left(\frac{q^2}{p^2}\right)^{-(N+1)/2}
               \left(\frac{q^2}{p^2}\right)^{-\gamma_L(N,a_s)/2}~,
\label{eqWV}
\end{equation}
where $a_s = \lambda_6^2/(4\pi)^3 = \sf const.$
The anomalous dimension $\gamma_L(N,a_s)$ is given by
\begin{equation}
\gamma_L(N,a_s) = \sqrt{(N+2)^2+1 -2 \sqrt{(N+2)^2 + 4a_s}}-(N+1)~.
\label{eqGAL}
\end{equation}
Note that $\gamma_L(N,a_s)$ possesses
 {\it no} poles but only branch cuts for
$N~\epsilon~{\rm\bf C}$. The anomalous dimension $\gamma_L$ covers all
conformal contributions in leading order. If one expands this quantity
it yields in first order in $a_s$ the {\it complete} leading order 
anomalous dimension $\gamma_{SS}^{(0)}(N)$,
up to an eventual term due to
4--momentum conservation  which is easily imposed,
\begin{equation}
\gamma_{SS}^{(0)}(N) = - \frac{2}{(N+1)(N+2)}+\frac{1}{6}~.
\end{equation}
Furthermore, all the fixed--order
leading poles at $N=-1$ are resummed in this
representation. This has been verified by an explicit calculation up to 
3--loop order~\cite{BVN}. The complete NLO fixed order anomalous 
dimension reads~\cite{BVN}
\begin{eqnarray}
\gamma_{SS}^{(1)}(N) &=&
 - \frac{1}{6} \frac{22 +111N+211N^2+138N^3 +28N^4}
{(N+1)^3(N+2)^3} + \frac{5}{3}\frac{S_1(N)}{(N+1)(N+2)} \nonumber\\
& &
- \frac{1}{2} \left[1+(-1)^N\right] \frac{2}{(N+1)^2(N+2)^2} 
+\frac{13}{216}~.
\end{eqnarray}

One may now derive from Eq.~(\ref{eqGAL}) the small--$x$ resummed
anomalous dimension, covering the fixed--order leading pole
contributions only
\begin{equation}
\gamma_L^{N \rightarrow -1}(N,a_s) = (N+1) \left[\sqrt{1- \frac{4a_s}
{(N+1)^2}} -1 \right]~,
\label{eqGAL1}
\end{equation}
which again contains {\it no} poles for all $N~\epsilon~{\rm\bf C}$.
This quantity corresponds to the LO BFKL anomalous dimension in QCD.

\begin{figure}[h]
\begin{center}
\includegraphics[width=.5\textwidth]{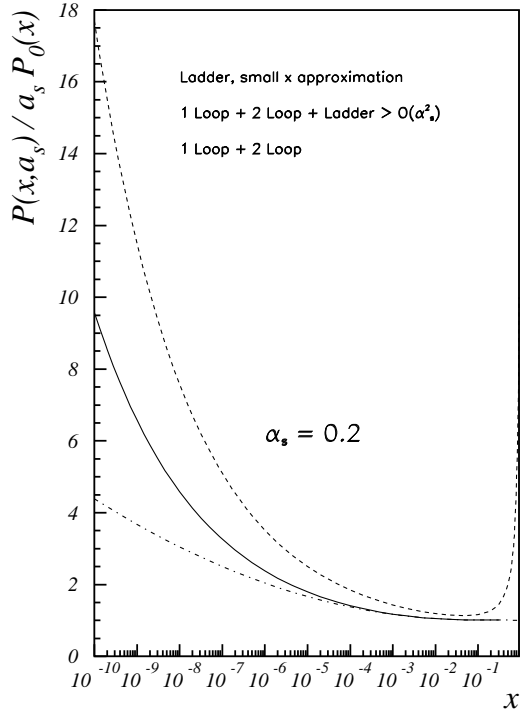}
\end{center}

\caption[]{Fixed-order and resummed splitting functions $P(x,a_s)$,
normalized to $a_sP_{SS}^{(0)}(x)$, for $a_s = 0.2$. Dash-dotted line~:
$P = a_s P_{SS}^{(0)} + a_s^2 P_{SS}^{(1)}$; solid line~: $P = P_L$;
dashed line~: $P = P_L^{x \rightarrow 0}$,~Ref.~\cite{BVN}.}
\label{BL-figphi}
\end{figure}

In deriving
(\ref{eqGAL1}) one obtains as well the respective resummed subleading
terms. As was shown in Ref.~\cite{BVN} the weight coefficient of these
terms are of alternating sign with growing coefficients, which indicates
already that
the resummation of the leading pole terms ($N=-1$) does not yield
the dominant contribution. This is expected, since neither 
$\gamma_L(N,a_s)$ nor $\gamma_L^{N \rightarrow -1}(N,a_s)$ have a pole
singularity -- as is
also the case in the LO BFKL resummation, where a
similar behaviour might be expected.   Fig.~\ref{BL-figphi}
shows the behaviour of the
respective splitting functions after the Mellin transform to $x$--space,
normalized to the leading order splitting function $P_0(x) = 2x(1-x)$.
$P_L^{x \rightarrow 0}(x,a_s)$ is
nowhere dominant and departs encreasingly
from the complete solution $P_L(x,a_s)$ as $x\rightarrow 0$.
\section{Conclusions}
%
As in the case of the fixed order calculations the renormalization group
equation, through which the factorization of the mass singularities is
described, implies the evolution equations for the parton densities
including the resummation of the small $x$ terms. Due to the Mellin
convolution between the respective evolution kernels and the extended
input distributions the detailed knowledge of the kernels at medium $x$
is as important. This is particularly the case for input densities
with a large rise towards
small $x$, as the gluon distribution. Less singular
contributions to the evolution kernels turn out to have a sizable
impact onto the scaling violations. In the example of $\phi^3_6$ theory
these contributions were calculated for 
the leading order resummation
and turn out to be even more important than the leading pole terms
($N = -1$). The reason for this behaviour is that the resummed
anomalous dimension, as also the resummed $(N = -1)$--fixed--order
pole contribution, possess {\it no poles} anymore. This is also the
case for the leading order resummed BFKL anomalous dimension and the
known resummed NLO contributions.

In
a quantitative description of the scaling violation of
structure functions             the conservation laws as Fermion number
conservation in the non--singlet case and energy--momentum conservation
have to be obeyed.
These integral relations imply strong relations between the small $x$
and medium $x$ contributions also for the resummed evolution kernels.
A study of the known fixed--order results in leading and 
next--to--leading order shows furthermore that the evolution kernels,
if approximated in a leading pole representation,  require to
take into account at least four orders which is likely to be the case
for the small $x$ resummed terms as well.
The conformal part of the known terms of the small $x$ resummations
behaves stable but is not necessarily dominant.

An important future check of the small $x$ resummed calculations
is
their prediction of the leading and next--to--leading order small $x$
contributions to the
3--loop anomalous dimensions, which are
yet unknown.
The intimate interplay between small and medium $x$ effects requires
to continue consistent calculations of the anomalous dimensions and
coefficient functions to even higher order and to compare these results
with the scaling violations measured by experiment.

\vspace{1mm}
\noindent
{\bf Acknowledgement.}~~For discussions I would like to thank A.~Vogt,
V.~Ravindran and W.L. van Neerven. This work was supported in part
by EC contract FMRX--CT98--0194.



%


\begin{thebibliography}{99}
\addcontentsline{toc}{section}{References}
%
\bibitem{GLR}
L.V. Gribov, E.M. Levin, and M.G. Ryskin: 
Nucl. Phys. \textbf{B188}, 555 (1983) \\
A. Mueller and J. Qiu: Nucl. Phys. \textbf{B268}, 427 (1986)
%
\bibitem{DLA}
A. De Rujula, S.L. Glashow, H.D. Politzer, S.B. Treiman, F. Wilczek,
and A. Zee: Phys, Rev. \textbf{D10}, 1649 (1974)\\
T. De Grand: Nucl. Phys. \textbf{B151}, 485 (1979)\\
J.P. Ralston and D.W. McKay, in: \emph{Physics Simulations at High
Energies}, ed. by V.~Barger (World Scientific, Singapore, 1987)\\
J. Bl\"umlein: Surv. High Energy Phys. \textbf{7}, 161 (1994)\\
R.D. Ball and S. Forte: Phys. Lett. \textbf{B336}, 77 (1994)
%
\bibitem{BV5}
J. Bl\"umlein and A. Vogt: Phys. Rev. \textbf{\bf D58}, 014020 (1998)
(1997)
%
\bibitem{GLR1}
J. Bartels, J. Bl\"umlein, and G. Schuler: Z. Phys. \textbf{C50}, 91
(1991)\\
J. Collins and J. Kwiecinski: Nucl. Phys. \textbf{B335}, 89 (1990) \\
M. Altmann, M. Gl\"uck, and E. Reya:  Phys. Lett. \textbf{B285}, 359
(1992)
%
\bibitem{BAR}
J. Bartels: Phys. Lett. \textbf{B298}, 204  (1993)
%
\bibitem{BFKL}
L.N. Lipatov: Sov. J. Nucl. Phys. \textbf{23}, 338 (1976)\\
E.A. Kuraev, L.N. Lipatov, and V.S. Fadin: Sov. Phys. JETP
\textbf{45}, 199 (1977) \\
I.I. Balitskii and L.N. Lipatov: Sov. J. Nucl. Phys. \textbf{28}, 822
(1978)\\
M. Ciafaloni: Nucl. Phys. \textbf{B296}, 49 (1988)
%
\bibitem{MAR}
G. Marchesini, in: \emph{QCD at 200 TeV}, ed. by L. Ciffarelli and 
Yu.L. Dokshitser, (Plenum Press, New York, 1992) pp.~183 and references
therein
%
\bibitem{JAR}
T. Jaroszewicz: Phys. Lett. \textbf{B116}, 291 (1982)
%
\bibitem{KL}
R. Kirschner and L.N. Lipatov: Nucl. Phys. \textbf{B213}, 122 (1983)
%
\bibitem{BV1}
J. Bl\"umlein and A. Vogt: Phys. Lett. \textbf{B370}, 149 (1996)
%
\bibitem{BER2}
J. Bartels, B.I. Ermolaev, and M.G. Ryskin: Z. Phys. \textbf{C76}, 241
(1997) 
%
\bibitem{BV2}
J. Bl\"umlein and A. Vogt: Acta Phys. Pol. \textbf{B27}, 1309 (1996)
%
\bibitem{BVR2}
J. Bl\"umlein, S. Riemersma, and A. Vogt: 
Eur. Phys. J. \textbf{C1}, 255 (1998)
%
\bibitem{CH}
S. Catani and F. Hautmann: Nucl. Phys. \textbf{B427}, 475 (1994)
%
\bibitem{FL}
V.S. Fadin and L.N. Lipatov: Phys. Lett. \textbf{B429}, 127 (1998)\\
V.S. Fadin: Preprint BUDKERINP-98/55; {\tt hep-ph/9807528}
%
\bibitem{CC2}
G. Camici and M.Ciafaloni: Nucl. Phys. \textbf{B496}, 305 (1997) \\
G. Camici and M.Ciafaloni: Phys. Lett. \textbf{B430}, 349 (1998)
%
\bibitem{CIA}
M. Ciafaloni: Phys. Lett. \textbf{B356}, 74 (1995)
%
\bibitem{DURH}
J. Bl\"umlein: J. Phys. \textbf{G19}, 1623 (1993)
%
\bibitem{WIL}
W.L.van Neerven: Talk, DESY Theory Workshop Sept. 1993
%
\bibitem{MW}
B.M. McCoy and T.T. Wu:  Phys. Lett. \textbf{B71}, 97 (1977)
%
\bibitem{BV3}
J. Bl\"umlein and A. Vogt: Phys. Lett. \textbf{B386}, 350 (1996)
%
\bibitem{EHW}
R.K. Ellis, F. Hautmann, and B. Webber: Phys. Lett. \textbf{B348}, 582
(1995)
%
\bibitem{BVR1}
J. Bl\"umlein, S. Riemersma, and A. Vogt: 
Nucl. Phys. \textbf{\bf B} (Proc. Suppl.)
\textbf{51C}, 30 (1996); Acta Phys. Pol. \textbf{B28}, 577 (1997)
%
\bibitem{BV4}
J. Bl\"umlein and A. Vogt: Phys. Rev. \textbf{\bf D57}, R1 (1998)
(1997)
%
\bibitem{BRNV} 
J. Bl\"umlein, V. Ravindran, W.L. van Neerven and A. Vogt:
`The Unpolarized Singlet Anomalous Dimension at Small x'. In:
\emph{DIS98, 6th International Workshop on Deep Inelastic Scattering
and QCD, Brussels, Belgium, April, 1998}, ed. by Gh. Coremans and 
R. Rosen (World Scientific, Singapore, 1998), pp. 211--216, 
{\tt hep-ph/9806368}
%
\bibitem{KOD}
Y. Kiyo, J. Kodaira, and H. Tochimura:  Z. Phys. \textbf{C74}, 631
(1997) 
%
\bibitem{EMR}
B.I. Ermolaev, S.I. Manyenkov, and M.G. Ryskin:
Z. Phys. \textbf{C69}, 259 (1996)
%
\bibitem{BER1}
J. Bartels, B.I. Ermolaev, and M.G. Ryskin: Z. Phys. \textbf{C70}, 273
(1996) 
%
\bibitem{BBN}
F. Berends, W.L. van Neerven, and G. Burgers: Nucl. Phys. \textbf{B297},
429 (1988); E:~\textbf{B304}, 921 (1988) 
%
\bibitem{JB95}
J. Bl\"umlein, `$k_{\perp}$ dependent parton densities in the
photon and proton.' In:  \emph{
       Proc. of the  XXX Renc. de Moriond}, Les Arcs, France, March
       1995, ed. J. Tran Than Van (Edition Frontieres, Paris, 1995), 
       pp. 191--197,
       {\tt hep-ph/9506446}  
%
\bibitem{KWI} 
J. Kwiecinski: Z. Phys. \textbf{C29}, 561 (1985)
%
\bibitem{FAD1}
V.S. Fadin, R. Fiore, and M. Kotsky: Phys. Lett. \textbf{B387}, 593 (1996)
%
\bibitem{BNR}
J. Bl\"umlein, V. Ravindran, and W.L. van Neerven:
Phys. Rev.  \textbf{D58}, 091502  (1998)
%
\bibitem{KOR}
I.A. Korchemskaya and G.P. Korchemsky: Phys. Lett. \textbf{B287}, 346
(1996)
%
\bibitem{EKL}
R.K. Ellis, E.M. Levin, and Z. Kunszt: Nucl. Phys. \textbf{B420}, 517
(1994); E:~\textbf{B433}, 498 (1995)
%
\bibitem{LOV}
C. Lovelace: Phys. Lett. \textbf{B55}, 187 (1975); Nucl. Phys. 
\textbf{B95}, 12 (1975)
%
\bibitem{BVN}
J. Bl\"umlein and W.L. van Neerven:
Phys. Lett. \textbf{B450}, 412  (1999)
\end{thebibliography}
\end{document}